# Orthogonal Evolution and Anticipation

by Hans-Rudolf Thomann

October 2008

Author address Im Altried 1h, 8051 Zürich, hr@thomannconsulting.ch

## CONTENTS

## Abstract

Quantum states evolving, for some $0 < p \leq \infty$, into a sequence of mutually orthogonal states at times $t + nT$ $(n = 0, \ldots, p - 1)$ exhibit an interesting physical effect. The analysis of the anticipation probabilities $p_n = \left|\left(q_{t+\frac{T}{2}}, q_{t+nT}\right)\right|^2$ shows, that for randomly chosen states $q$ measurements of the state at time $t + T/2$ reveal information about the state at time $t \pm nT$ $(n = 0, \ldots, p - 1)$, anticipating future states and reflecting past states with significant probability. For $p < \infty$ and any $0 < \delta < 1$, the probability to measure at time $t + \frac{T}{2}$ states from the set $\left\{q_{t+nT} \middle| \left|n - \frac{p}{2}\right| < \frac{p}{2}\delta\right\}$ exceeds $\sigma^2\delta$ with certainty as $p \to \infty$, where $\sigma^2$ is a constant independent from the period. For $p = \infty$ and any $N$, the probability to measure at time $t + \frac{T}{2}$ states from the set $\{q_{t\pm nT} | n > N\}$ is $\geq \sigma^2$. We characterize the spectrum, establish an analog to Planck's relation, define a random sampling scheme, analyze the resulting distribution of the anticipation probabilities and point out applications.

# 0 Introduction

Under a Hamiltonian $H$, a quantum state $q$ evolves into an orbit $q_t = U_t q$, where $U_t = e^{-\frac{iH}{\hbar}t}$. Due to the unitarity of $U_t$, the amplitude $d\hat{\mu}_q(T) = (q_t, q_{t+T}) = \int e^{-i\lambda T}\, d\mu_q(\lambda)$ is independent from $t$. If for some integer $0 < p \leq \infty$: $d\hat{\mu}_q(nT) = \delta_{n \bmod p}$ $(n \in \mathbb{Z})$, then $\{q_{t+nT} | n = 0, \dots, p-1\}$ is a set of mutually orthogonal states, for every $t$.

The subject of our study are states with this property, named *orthogonal evolution* of period $p$ and step size $T$, and the *anticipation amplitudes* $\alpha_n = d\hat{\mu}_q\big((n+1/2)T\big)$, i.e. the inner product of states at time $nT$ and $\frac{T}{2}$, as well as the *anticipation probabilities* $p_n = |\alpha_n|^2$. It turns out, that for randomly chosen states $q$, measurements of the state at time $\frac{T}{2}$ reveal information about the states $\pm nT$ $(n = 0, \dots, p-1)$. Future states are anticipated and past states reflected with significant probability. These probabilities and the strength of this effect, which we name Quantum-Mechanical *anticipation*, are quantified in this paper.

Chapter 1 shows that orthogonally evolving states have pure point or absolutely continuous spectrum $d\mu_q(\lambda)$, but neither mixed nor singular continuous spectrum. The fundamental upper bound $\hbar T^{-1} \leq \frac{2}{\pi} \inf_{\lambda_0} \langle |H| \rangle$ in terms of the spectral quantity $\langle |H| \rangle$ is established, which is an analog to Planck's relation. The spectral difference is defined in terms of the reduced spectrum and used to represent the anticipation amplitudes.

In chapter 2 we justify from the symmetries of the set of self-adjoint Hamiltonians the random sampling scheme underlying our analysis, derive the statistical distribution of the spectral difference as $q$ is randomly sampled from all orthogonally evolving states with period $p$ and step size $T$, identify the statistical quantities used in the subsequent chapters to assess anticipation strength and show that multiples of $p$ are unlikely to occur by chance.

In chapter 3 the anticipation effect is analyzed under point spectrum. The distribution of the $p_n$ and of various observables is derived for some model states as well as under random sampling. Theorem 1 states, that for any $0 < \delta < 1$, the probability to measure at time $\frac{T}{2}$ a state from the set $\left\{q_{nT} \,\middle|\, \left|n - \frac{p}{2}\right| < \frac{p}{2}\delta\right\}$ exceeds $\sigma^2\delta$ with certainty as $p \to \infty$, where $\sigma^2$ is a constant independent from the period. The operator $\langle \tilde{n}^r \rangle$ (defined in section 2.3) has expectation $\approx p^r$.

Chapter 4 determines anticipation under absolutely continuous spectrum. Theorem 2 states the constant lower bound $\sigma^2$ for the probability to measure at time $\frac{T}{2}$ a state from the set $\{q_{\pm nT} \,|\, n > N\}$, and infinite expectation for the operator $\langle \tilde{n}^r \rangle$. The statistics in the continuous case are the limiting values of those in the periodic case as $p \to \infty$.

In the final chapter we draw conclusions, discuss some open questions and point out applications. Anticipation in general type quantum evolution will be analyzed in a future paper.

# 1 Orthogonal Evolution

After introducing some terminology, we characterize the spectrum of orthogonally evolving states, derive necessary and sufficient conditions for their existence, establish a fundamental bound on the frequency and represent the anticipation amplitudes in terms of the reduced spectrum.

Definition 1

Let $H$ be a Hamiltonian, $q$ a state and $0 < p \leq \infty$ an integer. Throughout the paper we will assume that $\|q\|_2^2 = 1$.

a) $q$ is evolving under $H$ into $q_t = U_t q$, where $U_t = e^{-\frac{iH}{\hbar}t}$, has spectral measure $d\mu_q(\lambda)$.
b) $q$ has *orthogonal evolution* of period $p$ and step size $T$, iff
   $d\hat{\mu}_q(nT) = \delta_{n \bmod p} \ (n \in \mathbb{Z})$.
   Here $d\hat{\mu}_q(T) = (q_t, q_{t+T}) = \int e^{-i\lambda T/\hbar} d\mu_q(\lambda)$ is the Fourier transform of $d\mu_q(\lambda)$.
c) $\alpha_n = d\hat{\mu}_q\big((n + 1/2)T\big)$ is the *anticipation amplitude* of $q_{nT}$.
d) $p_n = |\alpha_n|^2$ is the *anticipation probability* of $q_{nT}$.
e) The *reduction of* $d\mu_q(\lambda)$ *modulo* $m$ is $d\nu_{q,m}(\kappa) \stackrel{\text{def}}{=} \sum_{\kappa = \frac{\lambda T}{\hbar} \bmod m} d\mu_q(\lambda)$.
f) The κ-cut of $d\mu_q(\lambda)$ modulo $m$ is the function $c_{\kappa,m}(n) = \frac{d\mu_q\left(nm + \frac{\kappa\hbar}{T}\right)}{d\nu_{q,m}(\kappa)} \ (n \in \mathbb{Z})$.
g) The *standard scale* is defined to make $\frac{T}{\hbar} = 1$.

## 1.1 Spectrum

Orthogonal evolution constrains the spectrum to either pure point or absolutely continuous, as stated by the following proposition.

Proposition 1: Characterization of the Spectrum (Orthogonal Evolution Constraint)

Let $q$ have orthogonal evolution with period $p$ and step size $T$. In standard scale,

a) If $p < \infty$, then $\nu_{q,2\pi}\left(\left\{\frac{2\pi k}{p}\right\}\right) = \frac{1}{p} \ (k = 0, \dots, p-1)$ is pure point.
b) If $p = \infty$, then $d\nu_{q,2\pi}(\kappa) = \frac{d\kappa}{2\pi} \ [0 \leq \kappa \leq 2\pi]$ is absolutely continuous.

Proof:

The spectrum of $H$ and $U_T$ are related by $U_t = e^{-\frac{iH}{\hbar}t}$. $U_T$ is a bounded operator with spectrum on the complex unit circle [RS80]. Its spectral measure is uniquely determined by $(f(U_T)q_0, q_0)$ $\big(f \in C^\infty(\mathbb{C})\big)$. These functions in turn are uniquely determined by the monomials $z^k$, which due to mutual orthogonality satisfy $(U_T^n q_0, q_0) = (q_{nT}, q_0) = \delta_{n \bmod p}$. Thus the spectral measure of $U_T$ is equally distributed, in the periodic case over the unit roots, in the infinite case over the whole complex unit circle. Singular continuous spectrum is excluded. Mixed spectrum cannot occur, as in the periodic case the initial state must recur, whereas in the aperiodic case $d\hat{\mu}_q(t) \in o(1)$ (see [YL96]).

The numerical values (in standard scale) follow from $\int_0^{2\pi} e^{-in\kappa} d\nu_{q,2\pi}(\kappa) = \delta_{n \bmod p}$ (see definition 1b) and the normalization of $q$. The integral becomes $p^{-1}\sum_{k=0}^{p-1} e^{-2\pi i n k/p}$ and $\int_0^{2\pi} e^{-in\kappa} \frac{d\kappa}{2\pi}$ for $p < \infty$ and $p = \infty$, respectively, which is easily verified to yield the correct values.

q.e.d.

A more elaborate proof facing the various intricacies and the explicit determination of the Fourier expansion of a continuous function $f$ on $[0,2\pi]$ such that $d\nu_{q,2\pi}(\kappa) = \mathrm{D}^2\mathrm{f}$ will be given in a future paper.

Proposition 1 has obvious consequences for the existence of orthogonal evolution.

Corollary 1: Existence

Necessary and sufficient for the existence of a state $q$ evolving orthogonally with period $p$ under Hamiltonian $H$ is that either $p < \infty$ and the reduced point spectrum of $H$ contains $p$ equidistant points (such as the $X^2$ operator), or $p = \infty$ and $H$ has non-empty absolutely continuous spectrum (such as the $-\Delta$ operator).

Proof:

In either case states meeting proposition 1 are easily constructed.

### 1.2 Fundamental Frequency Bound

In terms of the spectral quantity $\langle|H|\rangle$ we derive here the fundamental frequency bound $\hbar T^{-1} \leq \frac{2}{\pi}\langle|H|\rangle$, an analog to Planck's $E = \hbar\nu$.

Definition 2

For any Hamiltonian $H$ and state $q$, $\langle|H|\rangle \stackrel{\text{def}}{=} \int|\lambda - \lambda_0|d\mu_q(\lambda)$.

Proposition 2

Let $q$ be any state evolving under Hamiltonian $H$. Then

a) $\langle|H|\rangle < \infty$ iff $|\langle H\rangle| < \infty$.
b) $\langle|H|\rangle > 0$ iff $q \neq 0$ and $q$ not an eigenstates with zero eigenvalue.
c) $|(q, q_t)| \geq |Re(q, q_t)| \geq 1 - \langle|H|\rangle\, t/\hbar$.
d) If $d\hat{\mu_q}(T) = 0$, then $T \geq \sup_{\lambda_0}(\hbar/\langle|H - \lambda_0|\rangle)$.
e) The supremum is attained when $\lambda_0$ equals the median of $\mu$, i.e. $\mu([\lambda_0, \infty]) = \mu([-\infty, \lambda_0])$.
f) $\inf_{\lambda_0}\langle|H - \lambda_0|\rangle = 0$ iff $q$ is an eigenstate.

Proof:

a) By elementary measure theory, $\langle H\rangle = \int \lambda d\mu_q(\lambda)$ exists iff $\int_0^\infty \lambda d\mu_q(\lambda) < \infty$ and $\int_{-\infty}^0 \lambda d\mu_q(\lambda) < \infty$, i.e. iff $\langle|H|\rangle < \infty$.
b) As $\langle|H|\rangle$ is a sum of two non-negative quantities, it vanishes only, if $d\mu_q(\lambda) = 0$ or $supp\ \mu_q(\lambda) = \{0\}$
c) By the Schrödinger equation and the triangle inequality,
$$\left|\frac{d}{dt}\|q_t - q\|_2^2\right| = \left|\left(-i\frac{H}{\hbar}q_t, q\right) + \left(q, -i\frac{H}{\hbar}q_t\right)\right| = \left|2\int\frac{\lambda}{\hbar}\sin\frac{\lambda t}{\hbar}\,d\mu_q(\lambda)\right| \leq \frac{2}{\hbar}\langle|H|\rangle.$$
Therefore $\|q_t - q\|_2^2 \leq \frac{2}{\hbar}\langle|H|\rangle t$. The result follows from $Re(q, q_t) = 1 - \|q_t - q\|_2^2/2$
d) By orthogonality, $\|q_T - q\|_2^2 = 2$. As $T$ is the same for $H - \lambda_0$, the claimed result follows.

e) From $\frac{\partial y}{\partial \lambda_0}\langle|H-\lambda_0|\rangle = -\int_0^\infty d\mu_q(\lambda+\lambda_0) + \int_0^\infty d\mu_q(\lambda_0-\lambda) = 0.$
f) Elementary.

q.e.d.

2c) yields a lower bound on the modulus of the anticipation amplitude at $t = \frac{T}{2}$ . shows that at given $\langle|H|\rangle$ the frequency of the orthogonal evolution cannot be arbitrarily high, or equivalently the step size arbitrarily small. A bound for the passage time based on $\langle H^2\rangle$ has been established by Fleming [Fl73] and discussed by various authors. The bound given in 2d) for the passage time can be strengthened to a bound for the step size of orthogonal evolution:

Corollary 2: Fundamental Frequency Bound

The frequency of an orthogonal evolution is bounded by

$$\hbar T^{-1} \leq \frac{2}{\pi}\langle|H|\rangle.$$

Proof: Proposition 2d) and 2e) imply $\hbar T^{-1} \leq \langle|H-\lambda_0|\rangle$. We conclude from proposition 1 that among all Hamiltonians providing orthogonal evolution those with spectrum evenly distributed in $[-\pi,\pi]$ minimize $\langle|H|\rangle$, which is easily found to be $\frac{\pi}{2}$.

q.e.d.

### 1.3 Representation of Anticipation Amplitudes

From now on all spectra are in standard scale. In proposition 1 we have determined the reduced spectrum $d\nu_{q,2\pi}(\kappa)$. Using a notation for the reduced spectrum on the doubled interval $[0,4\pi]$ we establish a representation of $\alpha_n$ which we use throughout the paper.

Definition 3: Spectral difference

a) Let $p<\infty$. Then $x_{k,s} = \nu_{q,4\pi}\left(\left\{2\pi\left(s+\frac{k}{p}\right)\right\}\right)$ $(s=0,1; k=0,\dots,p-1)$.
$y_k = x_{k,0} - x_{k,1}$ is the spectral difference.
b) Let $p=\infty$. Then $x_{\kappa,s} = \frac{d}{d\kappa}\nu_{q,4\pi}(2\pi s+\kappa)$ $(s=0,1; 0\leq\kappa\leq 2\pi)$.
$y_\kappa = x_{\kappa,0} - x_{\kappa,1}$ is the spectral difference.

Proposition 3: Representation of anticipation amplitudes

a) If $p<\infty$, then $0 \leq x_{k,s} \leq \frac{1}{p}$, $-\frac{1}{p} \leq y_k \leq \frac{1}{p}$, $x_{k,0}+x_{k,1} = \frac{1}{p}$ .
b) If $p=\infty$, then $0 \leq x_{\kappa,s} \leq \frac{1}{2\pi}, -\frac{1}{2\pi} \leq y_\kappa \leq \frac{1}{2\pi}$, $x_{\kappa,0} + x_{\kappa,1} = \frac{1}{2\pi}$.
c) If $p<\infty$, then $\alpha_n = \sum_{k=0}^{p-1} y_k\, e^{-2\pi i\left(n-\frac{1}{2}\right)k/p}$ $(n\ integer)$ .
d) If $p=\infty$, then $\alpha_n = \int_0^{2\pi} y_\kappa e^{-i\left(n-\frac{1}{2}\right)\kappa}\frac{d\kappa}{2\pi}$ $(n\ integer)$.

Proof:

a) From proposition 1a) and definition 3a).
b) From proposition 1b) and definition 3b).
c) By definition 1c) and 3a),

$$\alpha_n = \sum_{k=0}^{p-1} x_{k,0} e^{-2\pi i\left(n-\frac{1}{2}\right)k/p} + x_{k,1} e^{-2\pi i\left(n-\frac{1}{2}\right)(1+k/p)}$$

$$= \sum_{k=0}^{p-1} \left(x_{k,0} - x_{k,1}\right) e^{-2\pi i\left(n-\frac{1}{2}\right)k/p}.$$

The claim follows from definition 3a).

d) By definition 1c) and 3b),

$$\alpha_n = \int_0^{2\pi} x_{\kappa,0} e^{-i\left(n-\frac{1}{2}\right)\kappa} \frac{d\kappa}{2\pi} + x_{\kappa,1} e^{-i\left(n-\frac{1}{2}\right)(2\pi+\kappa)} \frac{d\kappa}{2\pi}$$

$$= \int_0^{2\pi} \left(x_{\kappa,0} - x_{\kappa,1}\right) e^{-i\left(n-\frac{1}{2}\right)\kappa} \frac{d\kappa}{2\pi},$$

The claim follows from definition 3b).

q.e.d.

## 2 Random Sampling

Corollary 1 implies that orthogonally evolving states exist under infinitely many dynamics. In the remaining chapters we investigate, how likely such states are to exhibit anticipation of significant strength. To this end, we define in this chapter a random sampling scheme which will underlie all our subsequent analyses. We justify this scheme by the symmetry properties of the set of self-adjoint Hamiltonians. Under this scheme we derive the statistical distribution of the spectral difference, which determines the anticipation amplitudes. We define the set of statistical quantities used in the subsequent chapters to assess anticipation strength, and finally show that multiples of the period are unlikely to occur by chance.

### 2.1 Sampling Scheme

In order to investigate the likelihood of physical states to exhibit anticipation of a certain strength we must consider the set of all orthogonally evolving states and the way such states arise. We focus here on a laboratory situation, where the experimentalist prepares an experimental setup and an orthogonally evolving target, and then studies its evolution under the laws of nature. The setup is modeled by a Hamiltonian and the target by a physical state.

What Hamiltonians and states are preparable depends on the experimentalist's skills and capabilities and the constituents available in nature. Abstracting from these practical limitations, we only constrain the experimentalist to prepare self-adjoint Hamiltonians and states exhibiting orthogonal evolution, and analyze the outcome if these are randomly chosen, letting freely fluctuate all degrees of freedom except those fixed by the orthogonal evolution constraint (proposition 1).

We are going to define a sampling scheme, where, for a given step size and period, first a self-adjoint Hamiltonian is chosen from the set of all self-adjoint Hamiltonians admitting orthogonal evolution, and then a state is chosen from the set of states orthogonally evolving under that Hamiltonian.

Proposition 3 shows that orthogonal evolution and anticipation are completely determined by the spectral properties of the Hamiltonians and states involved, more specifically by the spectral difference. Thus, for our purpose, only constraints on and degrees of freedom of spectra need consideration.

Inverse theory tells us that, for any bounded real measure $\mu$ and any positive integer $d$, there are uncountably many self-adjoint operators $H$ and states $q \in L^2\left(\mathbb{R}^d\right)$ with spectral measure $\mu$. This holds as well for at least one Schrödinger operator $-\Delta + V$ with self-adjoint extension $H$. This enables us to define the sampling scheme in terms of probability spaces of spectral measures.

The core feature of spectral measures of orthogonally evolving states are the $\kappa$-cuts modulo $2\pi$ (see definition 1f), related to the spectral measure $\mu$ by $d\mu(2\pi n + \kappa) = c_{\kappa,2\pi}(n) d\nu_{q,2\pi}(\kappa)$. The quantities $x_{k,s}$ from definition 3 are $x_{k,s} = \sum c_{\kappa,2\pi}(2n + s) d\nu_{q,2\pi}(\kappa)$.

Whereas the reduced spectral measure is fixed by the orthogonal evolution constraint, the $\kappa$-cuts are arbitrary doubly-infinite sequences with non-negative values, summing up to 1.

Let's firstly consider their supports $supp\, c_{\kappa,2\pi}$, i.e. the patterns of points $n \in \mathbb{Z}$ where $c_{\kappa,2\pi}(n) \neq 0$. This pattern is a property of the Hamiltonian, whereas the size of $c_{\kappa,2\pi}(n)$ is a determined by the state.

Self-adjointness allows for any such pattern at any index $\kappa$, and to choose patterns at different indices $\kappa$ independently. This holds for point spectrum as well as for continuous spectrum, with the only constraint that for fixed $n$ the dependence from $\kappa$ must be absolutely continuous. While the following arguments will be given for point spectrum, they are easily extended to continuous spectrum as the limiting case.

This symmetry of the set of self-adjoint Hamiltonians allows us to require the sampling scheme to make the $supp\, c_{\kappa,2\pi}$ a family of independent identically distributed random variables. As these symmetries hold as well when constraining the Hamiltonians to admit orthogonal evolution, we demand the same property in this case.

Given any Hamiltonian, the value of $c_{\kappa,2\pi}(n)$ on $supp\, c_{\kappa,2\pi}$ depends only on the state $q$. When randomly sampling states $q \in L^2(\mathbb{R}^d)$ meeting the orthogonal evolution constraint, then $c_{\kappa,2\pi}$ can attain any non-negative function $f \in l^1(supp\, c_{\kappa,2\pi})$ meeting this constraint, independently for any index $\kappa$, because orthogonal components of these states belonging to different indices can be sampled separately in the discrete case. As by the preceding requirement the supports are i.i.d., we impose the same requirement on the sampling distribution of the $\kappa$-cuts.

As all distributions as well as the reduced spectrum are index-independent, we will omit indices in the following. Let $\theta_p$ be sampling distribution of the $\kappa$-cuts $c_{\kappa,2\pi}$ at period $p$.

Notice that $\theta_p$ is *not* obtained by conditioning from another, unconditional distribution. The orthogonal evolution constraint is met by normalization of the $\kappa$-cuts, not by conditioning of some distribution. Practical sampling procedures will be designed to meet the constraint approximately, but not by selection of "good" states from unconstraint samples. Only the latter procedure would yield a conditional distribution, whereas the real procedure is sufficiently modeled by a procedure where orthogonal components are sampled and normalized independently (index by index).

The period $p$ only determines the orthogonal evolution constraint. As this constraint is met by normalization, it only enters into $\theta_p$ as a scaling factor: $\theta_p(c_{\kappa,2\pi}) = \theta_q\left(\frac{p}{q} c_{\kappa,2\pi}\right)$ for any $p, q$ and $c_{\kappa,2\pi}$. As our reasoning did not depend otherwise from the period $p$, we finally postulate the sampling distribution of $\kappa$-cuts to be independent from $p$, up to the orthogonal evolution constraint and the scaling factor.

Our requirements are summarized in the following definition.

<u>Definition 4: Random Samples</u>

A family of sets $S_p$ of measures $\mu$ is a family of random samples of measures of orthogonally evolving states with period $p \leq \infty$ iff the following holds for all $p$:

a) The reduced spectrum $\nu_{q,2\pi}$ meets proposition 1 above.
b) For $p < \infty$, the $\kappa$-cut-valued random variables $c_\kappa$ are i.i.d. under probability measure $\theta_p(c) = \theta_1(pc)$ independent from $\kappa$.
c) For $p = \infty$, the $\kappa$-cut-valued random variables $c_\kappa$ are i.i.d. under probability measure $\theta_\infty(c) = \theta_1(2\pi c)$ independent from $\kappa$.

From the preceding arguments and definition 4 follows:

Corollary 3: Existence
The class of self-adjoint Hamiltonians admits for sampling schemes yielding random samples as defined above.

Further evidence justifying our sampling scheme will be provided by the main results and discussed in the final chapter.

## 2.2 Distribution of the Spectral Difference

Random sampling according to the above scheme yields a probability distribution of the spectral difference function $y_k$. It is no surprise that these are again i.i.d..

Proposition 4
Consider the sampling distribution as $q$ is randomly sampled under the sampling scheme of definition 4. Then for some distribution function $F(z)$ on $[-1,+1]$ the following holds:

a) If $p < \infty$, then $y_k$ $(k = 0, \dots, p-1)$ is a sequence of i.i.d. random variables with distribution $F(py)$ with density $pf(py)$ for $\left(-\frac{1}{p} \le y \le \frac{1}{p}\right)$.
b) If $p = \infty$, then $y_\kappa$ $(0 < \kappa \le 2\pi)$ is a family of i.i.d. random variables with distribution $F(2\pi y)$ with density $2\pi f(2\pi y)$ for $\left(-\frac{1}{2\pi} \le y \le \frac{1}{2\pi}\right)$.
c) If the random sampling scheme is invariant under translations of the energy scale, then the density $f$ is an even function.

Proof:

a) For $p < \infty$ the distribution of $x_{k,s}$ is $X_s(x) = \theta_1(\{c_{2\pi} | px \ge \sum c_{2\pi}(2n+s)\})$, from which $F$ is obtained. The scaling is due to proposition 1. The i.i.d. and period dependence are direct consequences of definition 4.
b) For $p = \infty$ the distribution of $x_{k,s}$ is $X_s(x) = \theta_1(\{c_{2\pi} | 2\pi x \ge \sum c_{2\pi}(2n+s)\})$, from which $F$ is obtained. The scaling is due to proposition 1. The i.i.d. and period dependence are direct consequences of definition 4.
c) Translations of the energy scale by $2\pi m$ effect a shift by $-m$ on the $\kappa$-cuts, transforming them to $c_{\kappa,2\pi}(n-m)$. If $m$ is odd, the sign of the spectral difference is toggled.

q.e.d.

## 2.3 Anticipation Statistics

We identify here the statistical quantities used to assess anticipation strength in the subsequent chapters, and recall some statistical terminology and basic facts.

The following anticipation statistics will be analyzed in the subsequent chapters, with $\tilde{n}$ as by definitions 5a and 5b below.

a) $p_n$, the probability to measure $q_{nT}$ at time $\frac{T}{2}$, and $p_N \stackrel{\text{def}}{=} P(\tilde{n} > N)$.

b) $p_{tot} = \sum p_n$, the probability to measure one of the states $q_{nT}$ at time $\frac{T}{2}$.
c) Observable $\tilde{n}^r$ $(r \geq 0)$, with value $\tilde{n}^r$ in state $q_{nT}$ $(n \geq 0)$ and expected value $\langle \tilde{n}^r \rangle \stackrel{\text{def}}{=} \sum \tilde{n}^r p_n$.

We perform the analysis for minimum and maximum anticipation model states as well as under biased and unbiased random sampling according to definition 4 above. In the sampling scenarios, we provide the sample mean and variance as well as stochastic asymptotics of the probabilities.

The following definitions will be used:

Definition 5

a) If $p < \infty$, $then\ \tilde{n} = \begin{cases} |n|\ mod\ p & (|n|\ mod\ p \leq\ (p+1)/2) \\ p+1-(|n|\ mod\ p) & (otherwise) \end{cases}$.
b) If $p = \infty$, $then\ \tilde{n} = |n|$.
c) $m_n = E(z^n)$ $(n = 0, 1, \dots)$ is the $n^{th}$ $moment$.
d) $\sigma^2(z) = E((z - m_1)^2)$ is the $variance$.
e) $z \in O_p(1)$ iff $\forall \varepsilon > 0\ \exists c\colon \limsup P(|z| > c) < \varepsilon$.
f) $z \in o_p(1)$ iff $\forall \varepsilon > 0\colon \lim P(|z| >\ \varepsilon) = 0$ .

$O_p$ and $o_p$ are Pratt's stochastic asymptotics [Pr59], the former expressing convergence in distribution, or stochastic boundedness, the latter stochastic convergence to zero [BFH75].

Recall the following facts:

Proposition 5

a) If the probability density is an even function, then $m_{2n+1} = 0$.
b) $E((cz)^n) = c^n m_n$.
c) $\sigma^2 = m_2 - m_1^2$.
d) $P\left(|z| > \left(E(|z|^r)\right)^{1/r} \Delta\right) < \frac{1}{\Delta^r}$ $(r > 0, \Delta > 0)$.
e) $P\left(\frac{|z - \overline{z}|}{\sigma} > \Delta\right) < \frac{1}{\Delta^2}$, where $\overline{z} = m_1$.
f) $z = \overline{z} + O_p(\sigma)$.
g) If $\sigma \in o(\overline{z})$ then $\frac{z}{\overline{z}} = 1 + o_p(1)$.

Proof:
Parts a-c) are elementary. d) is the Bienaymé-Chebychev inequality [TJ71], e) its special case $r = 2$, f) and g) direct consequences of e) and definition 5e-f).

## 2.4 Multiple Periods

Random sampling of orthogonally evolving states with period $p$ does admit states with period $np$ for some $n > 1$. Such states have $y_k = 0\ \forall k$, thus $\alpha_n = 0$. However, they are not likely to occur by chance in random samples.

Proposition 6

Let $q$ be a state with orthogonal evolution of period $p$.

a) $p < \infty$. For any $\varepsilon > 0$, let $n$ be the cardinality of the set $\{k \| p y_k| < \varepsilon, 0 \leq k < p\}$. Then $n$ has the Binomial distribution with parameter $p$ and $F(\varepsilon)$.
b) $p = \infty$. For any $\varepsilon > 0$, then the set $\{\kappa \| y_\kappa| < \varepsilon, 0 < \kappa \leq 2\pi\}$ has Lebesgue measure $F([-\varepsilon, +\varepsilon])$.

Proof:

a) Immediately from proposition 4.
b) From proposition 4, as the variables are i.i.d.

q.e.d.

This result rules out the occurrence by chance of period $np$ or, equivalently, step size $T/n$ when randomly sampling states with period $p$ and step size $T$. However it does not guarantee strong anticipation. This will be provided by the proof of constant probabilistic lower bounds on $p_N$ in the next two chapters.

## 3 Anticipation (Point Spectrum)

Here we determine, for model and random states with $p < \infty$, the values and distributions of the anticipation statistics identified and defined in section 2.3, and prove theorem 1.

### 3.1 Minimum Anticipation

We consider here the extremal case of constant $y_k$. Notice that these model states have zero occurrence probability.

<u>Proposition 7</u>

Let $py_k = y \ (k = 0, \dots, p-1)$ for some $-1 \le y \le 1$. Then

a) $p_n = \frac{y^2}{p^2 \sin^2 \pi\left(n-\frac{1}{2}\right)/p}$ with maximum value $\approx \frac{4y^2}{\pi^2}$ at $\tilde{n} = 0$ and $\tilde{n} = 1$, minimum value $\approx p^{-2}y^2$ at $\tilde{n} = \left[\frac{p}{2}\right]$ and $p_n \in O(\tilde{n}^{-2}) \ \left(\tilde{n} \to \left[\frac{p}{2}\right]\right)$.
b) $p_N \in O\left(\tilde{N}^{-1}\right) \ \left(\tilde{N} \to \left[\frac{p}{2}\right]\right)$.
c) $p_{tot} = y^2$.
d) $\langle \tilde{n}^1 \rangle = \ln p + O(1)$, $\langle \tilde{n}^r \rangle = O(p^{r-1}) \ (r > 1)$.

Proof:

a) By proposition 3c) and the premise

$$\alpha_n = p^{-1} y \sum_{k=0}^{p-1} e^{-2\pi i\left(n-\frac{1}{2}\right)k/p}$$

$$= \begin{cases} p^{-1} y \dfrac{e^{-2\pi i\left(n-\frac{1}{2}\right)} - 1}{e^{-2\pi i\left(n-\frac{1}{2}\right)/p} - 1} = \dfrac{-2p^{-1}y}{e^{-2\pi i\left(n-\frac{1}{2}\right)/p} - 1} = iy \dfrac{e^{-\pi i\left(n-\frac{1}{2}\right)/p}}{p \sin \pi\left(n-\frac{1}{2}\right)/p} & (2\tilde{n} - 1 \neq \mathrm{p}) \\ p^{-1} y & (\text{otherwise}) \end{cases}$$

The maxima, minima and asymptotics are elementary.
b) From a) by the Euler summation formula.
c) From Lemma 1 below.
d) From a) by the Euler summation formula.

q.e.d.

<u>Lemma 1</u>

$$\sum_{n=0}^{p-1} p^{-2} \sin^{-2} \pi\left(n - \frac{1}{2}\right)/p = 1$$

Proof:

Let $q$ be a superposition of exactly $p$ eigenfunctions. Then $Span(\{q_t | t \in \mathbb{R}\}) = Span(\{q_{kT} | k \in \mathbb{Z}\})$. Therefore $p_{tot} = \left\|q_{T/2}\right\|_2^2 = 1$. As in this case $|y| = |py_k| = 1$, the claimed result follows. The alternatively, direct summation, is easily achieved by adapting theorem 4.9a of [He88] to periodic functions.

### 3.2 Maximum Anticipation

We consider here the extremal case of alternating $y_k = (-1)^k y$. Notice that these model states have zero occurrence probability.

Proposition 8

Let $p$ be even, $py_k = (-1)^k y \ (k = 0, \dots, p-1)$ for some $-1 \le y \le 1$. Then

a) $p_n = \frac{y^2}{p^2 \cos^2 \pi\left(n-\frac{1}{2}\right)/p}$, which has maximum value $\approx \frac{4y^2}{\pi^2}$ at $\tilde{n} = \left\lceil \frac{p}{2} \right\rceil$, minimum value $\approx p^{-2}y^2$ at $\tilde{n} = 0$ and $\tilde{n} = 1$, and $p_n \in O(p^{-2}) \ (\tilde{n} \in o(p))$.

b) $p_N = p_{tot} - O(p^{-1}) \ \left(\tilde{N} \in o(p)\right)$.

c) $p_{tot} = y^2$.

d) $\langle \tilde{n}^r \rangle = O(p^r + p^{r-1} \ln p)$.

For odd $p$, this evolution degenerates to an orthogonal evolution with period $p$ and step size $T/2$.

Proof:

a) By proposition 3c) and the premise

$$\alpha_n = p^{-1} y \sum_{k=0}^{p-1} (-1)^k e^{-2\pi i \left(n-\frac{1}{2}\right)k/p}$$
$$= \begin{cases} p^{-1} y \frac{1-(-1)^p e^{-2\pi i\left(n-\frac{1}{2}\right)}}{1+e^{-2\pi i\left(n-\frac{1}{2}\right)/p}} = \frac{2p^{-1}y}{1+e^{-2\pi i\left(n-\frac{1}{2}\right)/p}} = y \frac{e^{-\pi i\left(n-\frac{1}{2}\right)/p}}{\mathrm{p} \cos \pi\left(n-\frac{1}{2}\right)/p} & (p \ even) \\ 0 & (p \ odd) \end{cases}.$$

The maxima, minima and asymptotics are elementary.

For odd $p$ and $y = \pm 1$, the support of $dv_{q,4\pi}(k)$ consists of $p$ equidistant points. Thus the evolution is of minimum anticipation type with period $p$ and step size $T/2$.

b) From a) by the Euler summation formula.

c) By Lemma 1.

d) From a) and the Binomial theorem one obtains $\langle \tilde{n}^r \rangle \approx \sum_{j,n} \left(\frac{p}{2}\right)^{r-j} (-n)^{j-2} \frac{r^{\underline{j}}}{j!}$, from which the result follows by the Euler summation formula. The notation $a^{\underline{b}}$ denotes the $b^{th}$ falling power of $a$.

q.e.d.

### 3.3 General Case

We consider now sampling distributions such that $y_k \ (k = 0, \dots, p-1)$ is a sequence of i.i.d. random variables with distribution $F_p(y) \ \left(-\frac{1}{p} \le y \le \frac{1}{p}\right)$ with moments $p^{-n} m_n$.

Proposition 9

a) $E(p_n) = p^{-1}\sigma^2 + \frac{m_1^2}{p^2 \sin^2 \pi\left(n-\frac{1}{2}\right)/p}$

$$= p^{-1}\sigma^2 + \begin{cases} O(p^{-2}) & (\tilde{n} \in \Omega(p)) \\ O(\tilde{n}^{-2}) & (\tilde{n} \in o(p)) \end{cases}.$$

$E(p_n)$ has maximum value $p^{-1}\sigma^2 + \frac{4}{\pi^2} m_1^2$ at $\tilde{n} = 0$ and $\tilde{n} = 1$.

$E(p_n)$ has minimum value $p^{-1}\sigma^2 + \frac{m_1^2}{p^2}$ at $\tilde{n} = \left\lceil \frac{p}{2} \right\rceil$.

b) $$p_n = E(p_n) + \begin{cases} = 0 & (m_n = m_1^n) \\ \in O_p(p^{-1/2}) & (2\tilde{n} - 1 = p) \\ \in O_p(p^{-1/2}\tilde{n}^{-1/2}) & (\tilde{n} \in o(p)) \\ \in O_p(p^{-1}) & (\tilde{n} \in \Omega(p), 2\tilde{n} - 1 \neq p) \\ \in O_p(p^{-1}) & (m_1 \in O(p^{-1/2})) \end{cases}.$$

c) $E(p_{tot}) = m_2$.

d) $$E(p_N) = (1 - 2p^{-1}\tilde{N})\sigma^2 + m_1^2 \begin{cases} 1 & (\tilde{N} = 0) \\ \frac{2}{\pi^2 \tilde{N}} & (\tilde{N} \in o(p), \tilde{N} \neq 0) \\ p^{-1}(1 - 2p^{-1}\tilde{N}) & (\text{otherwise}) \end{cases}.$$

e) $$p_N = E(p_N) + \begin{cases} = 0 & (m_n = m_1^n) \\ \in O_p(p^{-1/2}) & (otherwise) \end{cases}.$$

f) $E(\tilde{n}^r) = \frac{(p/2)^r}{r+1} m_2 + O(p^{r-1})$.

Proof:

a) From proposition 3c) and the premises, as detailed in the appendix.
b) From proposition 5f, where $Var(p_n)$ is calculated in the appendix.
c) See appendix.
d) See appendix.
e) From proposition 5f, where $Var(p_N)$ is calculated in the appendix.
f) From proposition 8a and the Euler summation formula.

q.e.d.

Proposition 9 shows that anticipation is fairly strong. Whereas the expected probability is independent of the period, the standard deviation is $O(p^{-1/2})$. We summarize these findings by

<u>Theorem 1: Anticipation (Point Spectrum)</u>

a) When randomly sampling with mean $m_1$ and $\sigma^2$, then, for any $0 < \delta < 1$, measurements at time $\frac{T}{2}$ yield a state from the set $\{q_{nT} | N < n < p + 1 - N\}$ with probability $p_N = \sigma^2 \delta + o_p(1)$ for $N = \frac{p}{2}(1 - \delta)$.
b) For any $0 < \delta < \sigma^2$ and any $N$, the probability $p_N > \delta$ with certainty as $p \to \infty$.
c) The operator $\langle \tilde{n}^r \rangle$ has expectation $O(p^r)$.

Proof:

a) By proposition 9b.
b) By a) and definition 5f).
c) By proposition 9f).

q.e.d.

Notice that for unbiased sampling $m_1 = 0$ and $\sigma^2 = m_2$, which by proposition 4c) is easily achievable.

## 4 Anticipation (Continuous Spectrum)

Here we determine, for model and random states with $p = \infty$, the values and distributions of the anticipation statistics identified and defined in section 2.3, and prove theorem 2.

### 4.1 Minimum Anticipation

We consider here the extremal case of constant $y_k$. Notice that these model states have zero occurrence probability.

Proposition 10

Let $y_\kappa = y \ (0 \le \kappa \le 2\pi)$ for some $-1 \le y \le 1$. Then

a) $p_n = y^2\pi^{-2}\left(n-\frac{1}{2}\right)^{-2}$ with maximum value $\approx \frac{4y^2}{\pi^2}$ at $\tilde{n} = 0$ and $\tilde{n} = 1$.
b) $p_N \in O(\tilde{N}^{-1})\ (\tilde{N} \to \infty)$.
c) $p_{tot} = y^2$.
d) $\langle \tilde{n}^r \rangle = \infty,\ (r \ge 1)$.

Proof:

a) By proposition 3d) and the premise
$$\alpha_n = \int_0^{2\pi} y e^{-i\left(n-\frac{1}{2}\right)\kappa} \frac{d\kappa}{2\pi} \ \ (n\ integer) = -iy\pi^{-1}\left(n-\frac{1}{2}\right)^{-1}$$
The maxima, minima and asymptotics are elementary.
b) From a) by the Euler summation formula.
c) By a well known result, first discovered by Euler. See also [He88].
d) From a).

q.e.d.

### 4.2 Maximum Anticipation

We consider here the extremal case of $y_k$ alternating between $y$ and $-y$ in $N$ equidistant intervals. Notice that these model states have zero occurrence probability.

Proposition 11

Let $y_\kappa = (-1)^j y \ \left(\frac{2\pi j}{M} \le \kappa < \frac{2\pi(j+1)}{M}, 0 \le j < M\right)$ for some even $0 < M < \infty$. Then

a) $p_n = \left[\frac{y}{\pi\left(n-\frac{1}{2}\right)}\tan\frac{\pi\left(n-\frac{1}{2}\right)}{M}\right]^2$ with minim value $\approx y^2M^{-2}$ at $n = 0$ and $n = 1$, and maximum value $\approx 4\pi^{-4}y^2$ at $n = \frac{M}{2}$.
b) $p_{tot} = y^2$
c) $p_N \begin{cases} = y^2 & (\mathrm{N} = 0) \\ \approx y^2\left(1 - O(\tilde{N}M^{-2})\right) & \left(\tilde{N} \in o(M)\right) \\ O(\tilde{N}^{-1}) & (\mathrm{N} \to \infty) \end{cases}$
d) $\langle \tilde{n}^r \rangle = \infty,\ (r \ge 1)$.

Proof:

a) By proposition 3d) and the premise

$$\alpha_n = \sum (-1)^j y \int_{\frac{2\pi j}{M}}^{\frac{2\pi(j+1)}{M}} e^{-i\left(n-\frac{1}{2}\right)\kappa} \frac{d\kappa}{2\pi}$$

$$= y \frac{e^{-2\pi i\left(n-\frac{1}{2}\right)/M} - 1}{-2\pi i\left(n-\frac{1}{2}\right)} \sum (-1)^j \, e^{-2\pi i\left(n-\frac{1}{2}\right)j/M}$$

$$= \mathrm{i}y \frac{e^{-2\pi i\left(n-\frac{1}{2}\right)/M} - 1}{2\pi\left(n-\frac{1}{2}\right)} \frac{(-1)^M e^{-2\pi i\left(n-\frac{1}{2}\right)} - 1}{e^{-2\pi i\left(n-\frac{1}{2}\right)/M} + 1}$$

$$= \frac{\mathrm{i}y}{\pi\left(n-\frac{1}{2}\right)} \tan\frac{\pi\left(n-\frac{1}{2}\right)}{M} \text{ for even } M.$$

For odd $M$ these model states have step size $\frac{T}{2}$.

b) Either by direct summation using theorem 4.9a of [He88], or using the fact that the functions $e^{-in\kappa}$ $(n \in \mathbb{Z})$ are an orthogonal basis of $\mathcal{L}^2([0,2\pi])$.
c) From a) and b).
d) By comparison with the minimum anticipation case.

q.e.d.

For odd $p$, this evolution degenerates to an orthogonal evolution with period $p$ and step size $T/2$.

### 4.3 General Case

We consider now sampling distributions such $y_\kappa$ $(0 < \kappa \le 2\pi)$ is a family of i.i.d. random variables with distribution $F(2\pi y)$ with moments $m_n$.

Proposition 12

a) $E(p_n) = \frac{m_1^2}{\pi^2\left(n-\frac{1}{2}\right)^2}$.

b) $E(p_{tot}) = m_2$.

c) $E(p_N) = m_2 - m_1^2 \begin{cases} 0 & (N = 0) \\ 1 - O(\tilde{N}^{-1}) & (N \to \infty) \end{cases}$.

d) $p_N = E(p_N)$ with probability 1.

e) $E(\tilde{n}^r) = \infty$, $(r \ge 1)$.

f) These values are the limits of the corresponding values as $p \to \infty$.

Proof:

a) $E(p_n) = \frac{1}{4\pi^2}\int_0^{2\pi}\int_0^{2\pi} E\big(y(\kappa)y(\kappa')\big)e^{-i\left(n-\frac{1}{2}\right)(\kappa-\kappa')}d\kappa\, d\kappa'$
$= \frac{m_1^2}{\pi^2\left(n-\frac{1}{2}\right)^2}.$

b) $p_{tot} = \sum_{n=-\infty}^{+\infty} \frac{1}{4\pi^2}\int_0^{2\pi}\int_0^{2\pi} y(\kappa)y(\kappa')e^{-i\left(n-\frac{1}{2}\right)(\kappa-\kappa')}d\kappa\, d\kappa'$
$= \lim_{m\to\infty} \frac{1}{4\pi^2}\int_0^{2\pi}\int_0^{2\pi} y(\kappa)y(\kappa')e^{-i\frac{\kappa-\kappa'}{2}}\frac{\sin m(\kappa-\kappa')}{\sin\frac{\kappa-\kappa'}{2}}d\kappa\, d\kappa',$

where taking the principal value is justified by the fact that $p_{tot} \le 1$. For further evaluation, we first notice that the functions $h_m(x) = \frac{\sin mx}{\sin\frac{x}{2}}$ $(-\pi \le x \le \pi)$ have a removable singularity at the origin and asymptotically vanishing integral for $x \in \Omega(m^{-1})$, whereas for $x \in \mathrm{o}(m^{-1})$ the approximations

$$h_m(x) = \frac{m\sin mx}{2\sin\frac{mx}{2}} + o(1) = m\cos m\frac{x}{2} + o(1)$$

hold. Approximating $\cos x$ in the interval $\left[\frac{-\pi}{2},\frac{\pi}{2}\right]$ by test functions and applying theorem 6.32 from [Ru73] yields a representation $\lim_{m\to\infty} m\cos m = 2\delta_x$ of Dirac's delta functional, such that

$$\lim_{m\to\infty}\frac{1}{2\pi}\int_{-\pi}^{+\pi} f(x)m\cos \mathrm{m}\frac{x}{2}dx = 2f(0).$$

The above arguments and the properties of $h_m$ imply that

$$\lim_{m\to\infty} e^{-i\frac{\kappa-\kappa'}{2}}\; h_m(x) = \delta_{\kappa-\kappa'} + \delta_{\kappa-\kappa'+\pi\ mod\ 2\pi}.$$

Therefore

$$p_{tot} = \frac{1}{4\pi^2}\int_0^{2\pi}\int_0^{2\pi} y(\kappa)y(\kappa')\,(\delta_{\kappa-\kappa'} + \delta_{\kappa-\kappa'+\pi\ mod\ 2\pi})d\kappa d\kappa'$$
$$= \frac{1}{2\pi}\int_0^{2\pi}\big(y^2(\kappa) + y(\kappa)y(\kappa+\pi\ mod\ 2\pi)\big)\,d\kappa,$$

from which the result follows by the independence property.

c) By a) and b) and the Euler summation formula.

d) Using the same methods as in the proof of b) above with the necessary adaptations we obtain
$E(p_N^2) =$
$$= E\left(\sum_{\tilde{n},\tilde{o}>N}\frac{1}{16\pi^4}\int_0^{2\pi} y(\kappa)y(\kappa')\,y(\kappa'')y(\kappa''')\,e^{-i\left(n-\frac{1}{2}\right)(\kappa-\kappa')-i\left(o-\frac{1}{2}\right)(\kappa''-\kappa''')}d\kappa d\kappa' d\kappa'' d\kappa'''\right)$$
$$= E\left(\left[\frac{1}{4\pi^2}\int_0^{2\pi} y(\kappa)y(\kappa')\left(\delta_{\kappa-\kappa'} - h_M(\kappa-\kappa')\right)d\kappa d\kappa'\right]^2\right)$$
$= E^2(p_N).$

Therefore $Var(p_N) = 0$. The result follows from proposition 5f.

e) By comparison with the minimum anticipation case.
f) By comparison with proposition 9.

q.e.d.

There is a seeming contradiction between 9a) and 9b), as $\sum p_n = m_1^2 \leq m_2 = p_{tot}$, which is due to the fact that $\lim_m \sum a_{n,m}$ does not necessarily equal $\sum \lim_m a_{n,m}$. The consequences are discussed in the final chapter.

Proposition 12 proves that anticipation strength in the continuous case is as strong as in the discrete case. The $p_N$ are of the same size, whereas the standard deviation is zero.

Theorem 2: Anticipation (Continuous Spectrum)

a) When randomly sampling with mean $m_1$ and standard deviation $\sigma$, then measurements at time $\frac{T}{2}$ yield a state from the set $\{q_{nT} | N < |n| < \infty\}$ with probability $p_N = m_2 - m_1^2 \left(1 - O\left(\tilde{N}^{-1}\right)\right)$ for all $0 \leq N < \infty$.
b) $p_N \geq m_2 - m_1^2$ with certainty for all $N$.
c) The operator $\langle \tilde{n}^r \rangle$ has infinite expectation.

Proof:

a) By proposition 12d.
b) By a) and definition 5f).
c) By proposition 12f).

q.e.d.

Notice that for unbiased sampling $m_1 = 0$ and $\sigma^2 = m_2$, which by proposition 4c) is easily achievable.

# 5 Conclusions

## 5.1 Summary

We have analyzed the statistical distribution of the anticipation probabilities $p_n$ and the cumulative probabilities $p_N$ when randomly sampling orthogonally evolving states with period $p$. The main findings are that, in the periodic case for any $0 < \delta < \sigma^2$ and $N = p\delta$, the probability $p_N > \delta$ with certainty as $p \to \infty$, and in the aperiodic case $p_N \geq \sigma^2$ with certainty for all $N$, where $\sigma^2$ is a constant independent from $p$. Somewhat surprisingly, the expected values of the single probabilities $p_n$ have the lower bound $p^{-1}\sigma^2$ independent of $n$, which is the main reason for the unexpected strength of anticipation.

## 5.2 Statistical Premises

Our analysis relies on a random sampling scheme justified by symmetries of the set of self-adjoint Hamiltonians described in section 2.1. Thus our results hold for random sampling from this set of Hamiltonians of which the set of physical Hamiltonians is a sub-set. Lacking an exhaustive characterization of the latter, we have based our analysis on the former.

Our results depend on three features of the distribution of the spectral difference (proposition 4): Independence from the index $\kappa$, statistical independence and independence from the period $p$. Let us discuss them one by one.

Independence from the index follows from the symmetries of the set of self-adjoint Hamiltonians. It is hard to imagine what physical effects should break these symmetries. Noticing that shifts of the energy scale shift the index $\kappa$, how should any asymmetry arise. We assert that independence from the index holds as well for the set of physical Hamiltonians.

Statistical independence of the $y_k$ is provided by separately sampling the orthogonal components of a state for different indices $k$. Though this may not be practical, it shows that the assumption of statistical independence is legitimate as a model of reality and approximation to experimental practice.

Independence from the index and statistical independence imply that the anticipation strength mainly depends on the variance $\sigma^2$, as established in chapters 3 and 4.

Independence from the period may indeed be broken by some law of nature, or by the particular techniques used by the experimentalist. But what will this effect? Will it decrease or increase the variance? How and why should a physical or a preparation effect decrease the variance of a single spectral difference $y_k$ below $p^{-2}\sigma^2$ as $p \to \infty$, which is the value we have assumed? The central limit theorem does not apply in this case, and conditioning has also been ruled out.

As precision is rather expected to decrease than to increase, we maintain our model as a good approximation to physical reality and assert that our findings are valid for physical Hamiltonians and practical realizations.

## 5.3 Features and Limitations

Orthogonally evolving states are strange objects, as they seem to flip back and forth between present and distant times every $\frac{T}{2}$ time. The observable $\langle(\tilde{n} - t)^r\rangle$ oscillates between $0$ at full steps and $O(p^r)$ at half steps. Measurements at $\frac{T}{2}$ may yield states at distant times.

However, this behavior is just due to coherence, not to any non-locality in space or time. Future measurements do not affect the state at time $\frac{T}{2}$, nor does measurement at that time affect the past state.

The oscillations may effect translations in space and energy transport. Their size is thus subject to energetic and relativistic limitations.

As an example, periodic motion under a Schrödinger Hamiltonian is accompanied by an oscillation of the kinetic and potential energy. Thus measurements of $\langle -\Delta \rangle$ at full and half steps will exhibit such oscillations, their amplitude being related to $\frac{T}{2}[V, -\Delta]$ by the mean value theorem of calculus.

In the continuous case, $\langle \tilde{n}^r \rangle$ is related to $\langle X^r \rangle_n$, as states with absolutely continuous spectrum are known to spread to infinity [KL99, KKL01,Si90]. Therefore $\langle X^r \rangle$ is infinite for orthogonally evolving states. $\langle (\tilde{n} - t)^r \rangle$ is not measureable at time $\frac{T}{2}$, because the state is not localized, whereas measurement equipment is. However, successive measurements of their presence at full and half steps will reveal a difference of relative size $\sigma^2$, thus proving the disappearance of the target.

### 5.4 Applications

Preparation of orthogonally evolving states followed by measurement after one half-step is a generic method for rapid state preparation.

This obviously applies to quantum computing, which is nothing but preparation of the result state from the initial state. It has been explained in [TH0801] that quantum computers simulating computations of deterministic Turing machines evolve orthogonally, and how the anticipation effect can be exploited to obtain the results of arbitrarily long computations at time $\frac{T}{2}$.

### 5.5 Open Problems

To our knowledge, orthogonal evolution and the associated anticipation effect have so far not been a topic of physical literature and experimental practice. One reason may be, that this effect is neither of interest nor easily observable in fast processes of short periods, which are predominant in microscopic systems.

This is different for quantum computers. Their first generation is likely to be of hybrid type, with a quantum memory and a classical finite state control and thus not exhibit anticipation. Anticipation will be achieved once not only the state but also the finite state control are in a coherent state, thus implementing a time-invariant Hamiltonian. We will outline this topic and its applications to hypercomputing in future papers.

The first step towards verification of anticipation will be to prepare quantum systems with long periods, and to prepare them such that they evolve orthogonally. Composite systems of $p$ quantum oscillators with suitable frequencies are candidate realizations. Once this has been done, the next task is the measurement of the predicted anticipation effect.

Possibly, nature is ahead of us once again, and is using anticipation to speed-up processes. It would thus be worthwhile to seek for physical phenomena arising due to anticipation.

An obvious question is whether anticipation is specific for orthogonal evolution. We will analyze anticipation in general type quantum evolution in a future paper demonstrating that anticipation is a universal phenomenon.

# 7 Appendix: Proofs of Proposition 8

## 7.1 Mean and Variance of $p_n$

The tilde operator is defined in definition 5a-b. Define

$$\pi_N = 1 - \frac{2\tilde{N}}{p}$$
$$\pi_n^{'} = 1 - np^{-1}$$

$$s_k \stackrel{\text{def}}{=} s_k(n) \stackrel{\text{def}}{=} p^{-1}e^{-2\pi i\left(n-\frac{1}{2}\right)k/p}, s_0 = |s_k| = p^{-1}.$$

$$S \stackrel{\text{def}}{=} S_n \stackrel{\text{def}}{=} \sum_{k=0}^{p-1} s_k = p^{-1}\frac{e^{-2\pi i\left(n-\frac{1}{2}\right)} - 1}{e^{-2\pi i\left(n-\frac{1}{2}\right)/p} - 1}$$
$$= \frac{-\mathrm{i}e^{\pi i\left(n-\frac{1}{2}\right)/p}}{p\sin\pi\left(n-\frac{1}{2}\right)/p}\begin{cases}\approx \frac{1}{\pi\tilde{n}} & (\tilde{n} \in o(p)) \\ \in O(p^{-1}) & (otherwise)\end{cases},$$

which satisfies

$$\overline{S_n} = S_{p+1-n}$$

and by Lemma 1

$$\textstyle\sum_n |S|^2 = 1 \ .$$

Using

$$T_n \stackrel{\text{def}}{=} p^{-2}\frac{e^{-2n\pi i} - 1}{e^{-2n\pi i/p} - 1} = \begin{cases}p^{-1} & (n = 0 \ mod \ p) \\ 0 & (otherwise)\end{cases}$$

$T \stackrel{\text{def}}{=} |T_{2n-1}|$ which for even $p$ is always zero,

we verify easily

$$\textstyle\sum s_j(n)s_k(o) = S_n S_o,$$

$$\textstyle\sum s_j(n)s_{-k}(o) = S_n\overline{S_o},$$

$$\sum s_j(n)s_j(o) = T_{n+o-1}$$

$$\sum s_j(n)s_{-j}(o) = T_{n-o}$$

$$\sum_{j\neq k} s_j(n)s_k(o) = S_n S_o - T_{n+o-1},$$

$$\textstyle\sum_{j\neq k} s_j(n)s_{-k}(o) = S_n\overline{S_o} - T_{n-o}.$$

Now by proposition 3c)

$$E(p_n) = E\left(\left|\sum_{k=0}^{p-1} y_k e^{-2\pi i\left(n-\frac{1}{2}\right)k/p}\right|^2\right) = p^{-2}\sum_k (m_2 - m_1^2) + \sum_{j,k} m_1^2 s_{j-k}$$

$$= p^{-1}(m_2 - m_1^2) + m_1^2 S^2.$$

$\sigma^2 = m_2 - m_1^2$ is the variance and thus non-negative.

$E(p_{tot}) = E(\sum_n p_n) = \sum_n E(p_n)$ and Lemma 1 immediately yield

$E(p_{tot}) = m_2.$

Again by proposition 3c)

$$E(p_n^2) = E\left(\left|\sum_{k=0}^{p-1} y_k e^{-2\pi i\left(n-\frac{1}{2}\right)k/p}\right|^4\right) = E\left(\sum_{j,k,l} y_j y_k y_l y_m s_j s_{-k} s_l s_{-m}\right),$$

which we evaluate by partitioning the index set and collecting the contributions of each sub-set:

| | | |
|---|---|---|
| a) | $j = k = l = m$ | $p^{-3}m_4$ |
| b) | $j \neq k = l = m$ | $p^{-2}(S^2 - p^{-1})m_1 m_3$ |
| | $k \neq j = l = m$ | |
| | $l \neq k = j = m$ | |
| | $m \neq k = l = j$ | |
| c) | $j = k \neq l = m$ | $p^{-4}p^2 m_2^2$ |
| | $j = m \neq k = l$ | |
| d) | $j = l \neq k = m$ | $(T^2 - p^{-3})m_2^2$ |
| e) | $j = k \neq l \neq m$ | $(p^{-1} - 2p^{-2})(S^2 - p^{-1})m_1^2 m_2$, |
| | | using $\{j = k \neq l, m\} - \{j = k \neq l = m\}$ |
| | | where the first bracket yields $p^{-2}\sum(S - s_j)\overline{(S - s_j)}$ |
| | | together $p^{-2}(p-2)S^2 + 2p^{-3} - p^{-2}$ |
| | $j \neq k \neq l = m$ | using $\{l = m \neq j, k\} - \{l = m \neq j = k\}$ |
| | $j = m \neq k \neq l$ | using $\{j = m \neq k, l\} - \{j = m \neq k = l\}$ |
| | $k = l \neq j \neq m$ | using $\{k = l \neq j, m\} - \{k = l \neq, j = m\}$ |
| f) | $j = l \neq k \neq m$ | $\left((T - 2p^{-2})S^2 + 2p^{-3} - T^2\right)m_1^2 m_2$, |
| | | using $\{j = l \neq k, m\} - \{j = l \neq k = m\}$ |
| | | where the first bracket yields |
| | | $\sum s_j^2 \left(S - s_{-j}\right)^2 = \sum s_j^2\left(S^2 - 2S s_{-j} + s_{-j}^2\right)$ |
| | | together $(T - 2p^{-2})S^2 + 2p^{-3} - T^2$ |
| | $k = m \neq j \neq l$ | using $\{k = m \neq j, l\} - \{k = m \neq j = l\}$ |
| g) | $j \neq k \neq l \neq m$ | $(S^4 + (8p^{-2} - 2T - 4p^{-1})S^2 + T^2 - 6p^{-3} + 2p^{-2})m_1^4$, |
| | | using $\{ j \neq k, l, m\} - \{j \neq k \neq l = m\} - \{j \neq k = l \neq m\} -$ $\{j \neq k = m \neq l\} - \{j \neq k = l = m\}$ |
| | | where the first bracket yields $\sum s_j \left(S - s_j\right)\left(S - s_{-j}\right)^2$ |
| | | $= \sum s_j S^3 - \left(s_j^2 + 2p^{-2}\right)S^2 + 3p^{-2} s_j S - p^{-4}$ |

$$= S^4 - (T + 2p^{-1} - 3p^{-2})S^2 - p^{-3}$$

Together

$$S^4 - (T + 2p^{-1} - 3p^{-2})S^2 - p^{-3} - 2(p^{-1} - 2p^{-2})(S^2 - p^{-1}) - \big((T - 2p^{-2})S^2 + 2p^{-3} - T^2\big) - p^{-2}(S^2 - p^{-1})$$
$$= S^4 - (2T + 4p^{-1} - 8p^{-2})S^2 + T^2 - 6p^{-3} + 2p^{-2}$$

This yields

$$\begin{aligned} E(\mathrm{p}_\mathrm{n}^2) = {} & p^{-3}m_4 + 4p^{-2}(S^2 - p^{-1})m_1m_3 + 2p^{-4}p^2m_2^2 + (T^2 - p^{-3})m_2^2 \\ & + 4(p^{-1} - 2p^{-2})(S^2 - p^{-1})m_1^2m_2 \\ & +2\big((T - 2p^{-2})S^2 + 2p^{-3} - T^2\big)m_1^2m_2 \\ & + (S^4 + (8p^{-2} - 2T - 4p^{-1})S^2 + T^2 - 6p^{-3} + 2p^{-2})m_1^4 \\ = {} & p^{-3}m_4 + 4p^{-2}(S^2 - p^{-1})m_1m_3 + (T^2 + 2p^{-2} - 3p^{-3})m_2^2 \\ & +\big((2T + 4p^{-1} - 12p^{-2})S^2 + 12p^{-3} - 4p^{-2} - 2T^2\big)m_1^2m_2 \\ & + (S^4 + (8p^{-2} - 2T - 4p^{-1})S^2 + T^2 - 6p^{-3} + 2p^{-2})m_1^4 \end{aligned}$$

The variance is

$$\begin{aligned} Var(\mathrm{p}_\mathrm{n}) = {} & E(p_n^2) - p^{-2}(m_2 - m_1^2)^2 - m_1^4S^4 - 2p^{-1}m_1^2(m_2 - m_1^2)S^2 \\ = {} & E(p_n^2) - p^{-2}m_2^2 + 2(p^{-2} - p^{-1}S^2)m_1^2m_2 + (2p^{-1}S^2 - \mathrm{p}^{-2} - S^4)m_1^4 \\ = {} & p^{-3}m_4 + 4p^{-2}(S^2 - p^{-1})m_1m_3 + (T^2 + p^{-2} - 3p^{-3})m_2^2 \\ & +\big((2T + 2p^{-1} - 12p^{-2})S^2 + 12p^{-3} - 2p^{-2} - 2T^2\big)m_1^2m_2 \\ & + \big((8p^{-2} - 2T - 2p^{-1})S^2 + T^2 - 6p^{-3} + p^{-2}\big)m_1^4 \end{aligned}$$

One easily verifies

$$Var(p_n)\begin{cases} = 0 & (m_n = m_1^n) \\ \in \mathrm{O}(p^{-1}) & (2\tilde{n} - 1 = p\ ) \\ \in \mathrm{O}(p^{-1}\tilde{n}^{-1}) & \big(\tilde{n} \in \mathrm{o}(p)\big) \\ \in \mathrm{O}(p^{-2}) & (\tilde{n} \in \Omega(p), 2\tilde{n} - 1 \neq p) \\ \in \mathrm{O}(p^{-2}) & \Big(m_1 \in \mathrm{O}\big(p^{-1/2}\big)\Big) \end{cases}$$

## 7.2 Mean and Variance of $p_N$

We define

$$U_N = \sum_{\tilde{n}>N} |S_n|^2 \approx \begin{cases} \dfrac{2}{\pi^2 \tilde{N}} & \left(\tilde{N} \in o(p)\right) \\ p^{-1}\pi_N & (\text{otherwise}) \end{cases}$$

and notice

$$\textstyle\sum_{\tilde{n},\tilde{o}>N} T_{n+o-1} S_n S_o = \sum_{\tilde{n},\tilde{o}>N} T_{n-o} S_n \bar{S}_o = p^{-1} U_N.$$

From the preceding results we find

$$E(p_N) = \sum_{\tilde{n}>N} E(p_n) = \pi_N (m_2 - m_1^2) + m_1^2 U_N$$

$$\approx \pi_N \sigma^2 + m_1^2 \begin{cases} 1 & \left(\tilde{N} = 0\right) \\ \dfrac{2}{\pi^2 \tilde{N}} & \left(\tilde{N} \in o(p), \tilde{N} \neq 0\right) \\ p^{-1}\pi_N & (otherwise) \end{cases}$$

$$E(p_N^2) = E\left(\left(\sum_{\tilde{n}>N} p_n\right)^2\right) = \sum_{\tilde{n},\tilde{o}>N} E(p_n p_o)$$

$$= E\left(\textstyle\sum_{\tilde{n},\tilde{o}>N} \sum_{j,k,l,m} y_j y_k y_l y_m s_j(n)\, s_{-k}(n) s_l(o) s_{-m}(o)\right),$$

which we evaluate as before by partitioning the index set and collecting the contributions of each sub-set:

a) $j = k = l = m$ — $p^{-1}\pi_N^2 m_4$

b) $j \neq k = l = m$ — $p^{-1}\pi_N (U_N - \pi_N) m_1 m_3$
   $k \neq j = l = m$
   $l \neq k = j = m$
   $m \neq k = l = j$

c) $j = k \neq l = m$ — $\pi_1^{'} \pi_N^2 m_2^2$

d) $j = m \neq k = l$ — $p^{-1}\pi_N (1 - \pi_N) m_2^2$
   which follows from
   $\sum |T_{n-o}|^2 - p^{-3} = p^{-1}\pi_N - p^{-1}\pi_N^2$

e) $j = l \neq k = m$ — $p^{-1}\pi_N (1 - \pi_N) m_2^2$

f) $j = k \neq l \neq m$ — $\pi_2^{'} \pi_N (U_N - \pi_N) m_1^2 m_2$
   using $\{j = k \neq l, m\} - \{j = k \neq l = m\}$
   where the first bracket yields
   $p^{-2} \sum \left(S_o - s_j(o)\right)\left(\overline{S_o} - s_{-j}(o)\right)$=$(p^{-1} - 2p^{-2})|S_o^2| + p^{-3}$
   summing up to
   $\pi_2^{'} \pi_N U_N + p^{-1}\pi_N^2$

   $j \neq k \neq l = m$ — using $\{l = m \neq j, k\} - \{l = m \neq j = k\}$

g) $j = m \neq k \neq l$ — $p^{-1}(1 - 2\pi_N)(U_N - \pi_N) m_1^2 m_2$

using $\{j = m \neq k, l\} - \{j = m \neq k = l\}$
where the first bracket yields

$$\sum s_j(n)s_{-j}(o)\left(\overline{S_n} - s_{-j}(n)\right)\left(S_o - s_j(o)\right)$$
$$= T_{n-o}\overline{S_n}S_o - p^{-2}|S_n|^2 - p^{-2}|S_o|^2 + p^{-3}$$

which sums up to
$p^{-1}U_N - 2p^{-1}\pi_N U_N + p^{-1}\pi_N^2 = p^{-1}U_N(1 - 2\pi_N) + p^{-1}\pi_N^2$

$k = l \neq j \neq m$ — using $\{k = l \neq j, m\} - \{k = l \neq j = m\}$

h) $j = l \neq k \neq m$ — $p^{-1}(1 - 2\pi_N)(U_N - \pi_N)m_1^2 m_2$,
using $\{j = l \neq k, m\} - \{j = l \neq k = m\}$
where the first bracket yields

$$\sum s_j(n)s_j(o)\left(\overline{S_n} - s_{-j}(\mathrm{n})\right)\left(\overline{S_o} - s_{-j}(\mathrm{o})\right)$$
$$= \sum_{\mathrm{n,o}>N} T_{n+o-1}\overline{S_n S_o} - p^{-2}|S_n^2| - p^{-2}|S_o^2| + p^{-3}$$

summing up to
$p^{-1}(U_N - 2\pi_N U_N) + p^{-1}\pi_N^2$

$k = m \neq j \neq l$ — using $\{k = m \neq j, l\} - \{k = m \neq j = l\}$

i) $j \neq k \neq l \neq m$ — $\Big(U_N^2 + U_N(8p^{-1}\pi_N - 4p^{-1} - 2\pi_N) + \pi_N(\pi_N + 2p^{-1} -$
*6p−1πNm14*

using $\{ j \neq k, l, m\} - \{j \neq k \neq l = m\} - \{j \neq k = l \neq m\} - \{j \neq k = m \neq l\} - \{j \neq k = l = m\}$
where the first bracket yields

$$\sum s_j(\mathrm{n})\left(\overline{S_n} - s_{-j}(\mathrm{n})\right)\left(S_o - s_j(\mathrm{o})\right)\left(\overline{S_o} - s_{-j}(\mathrm{o})\right)$$
$$= \sum |S_n^2||S_o^2| - \overline{S_n}S_o T_{n-o} - \overline{S_n S_o}T_{n+o-1} + p^{-2}|S_n^2| - p^{-1}|S_o^2|$$
$$+ p^{-2}|S_o^2| + p^{-2}|S_o^2| - p^{-3}$$
$$= U_N^2 - 2p^{-1}U_N + 3p^{-1}\pi_N U_N - \pi_N U_N - p^{-1}\pi_N^2$$
$$= U_N^2 - 2p^{-1}U_N + p^{-1}\pi_N(U_N - \pi_N) + 2p^{-1}\pi_N U_N - \pi_N U_N$$
$$= U_N^2 + 2p^{-1}U_N(\pi_N - 1) + p^{-1}\pi_N(U_N - \pi_N) - \pi_N U_N$$

together

$$U_N^2 + 2p^{-1}U_N(\pi_N - 1) + p^{-1}\pi_N(U_N - \pi_N) - \pi_N U_N$$
$$-\pi_2'\pi_N(U_N - \pi_N)$$
$$-p^{-1}(1 - 2\pi_N)(U_N - \pi_N)$$
$$-p^{-1}(1 - 2\pi_N)(U_N - \pi_N)$$
$$-p^{-1}\pi_N(U_N - \pi_N)$$
$$= U_N^2 + \left(-\pi_2'\pi_N - 2p^{-1}(1 - 2\pi_N)\right)(U_N - \pi_N) + 2p^{-1}U_N(\pi_N - 1) - \pi_N U_N$$
$$= U_N^2 + (U_N - \pi_N)(6p^{-1}\pi_N - \pi_N - 2p^{-1}) + 2p^{-1}U_N(\pi_N - 1) - \pi_N U_N$$
$$= U_N^2 + U_N(8p^{-1}\pi_N - 4p^{-1} - 2\pi_N) + \pi_N(\pi_N + 2p^{-1} - 6p^{-1}\pi_N)$$

Check: $m_k = m_1^k \rightarrow E(p_N^2) = m_1^4 U_N^2$

The variance is

$$\begin{aligned}
Var(p_N) &= E(p_N^2) - (\pi_N(m_2 - m_1^2) + m_1^2 U_N)^2 \\
&= E(p_N^2) - \pi_N^2 m_2^2 + 2\pi_N(\pi_N - U_N)m_1^2 m_2 - (\pi_N - U_N)^2 m_1^4 \\
&= p^{-1}\pi_N^2 m_4 + 4p^{-1}\pi_N(U_N - \pi_N)m_1 m_3 + (\pi_1'\pi_N^2 + 2p^{-1}\pi_N(1-\pi_N) - \pi_N^2)m_2^2 \\
&\quad + \big((2\pi_2'\pi_N + 4p^{-1}(1-2\pi_N) - 2\pi_N)(U_N - \pi_N)\big)m_1^2 m_2 \\
&\quad + (U_N^2 + U_N(8p^{-1}\pi_N - 4p^{-1} - 2\pi_N) + \pi_N(\pi_N + 2p^{-1} - 6p^{-1}\pi_N) \\
&\quad - \pi_N^2 + 2\pi_N U_N - U_N^2)m_1^4 \\
&= p^{-1}\pi_N^2 m_4 + 4p^{-1}\pi_N(U_N - \pi_N)m_1 m_3 + p^{-1}\pi_N(2 - 3\pi_N)m_2^2 \\
&\quad + 4p^{-1}(1-3\pi_N)(U_N - \pi_N)m_1^2 m_2 + 2p^{-1}\big(2U_N(2\pi_N - 1) + \pi_N(1 - 3\pi_N)\big)m_1^4
\end{aligned}$$

One easily verifies

$$Var(\mathrm{p_n}) \begin{cases} = 0 & (m_n = m_1^n) \\ \in O(p^{-1}) & (\text{otherwise}) \end{cases}$$

$$\begin{aligned}
&\pi_N^2 + \pi_N(U_N - \pi_N) + \pi_N(2 - 3\pi_N) \\
&\quad +4(1-2\pi_N)(U_N - \pi_N) + 2\big(2U_N(2\pi_N - 1) + \pi_N(1-3\pi_N)\big)
\end{aligned}$$